\documentclass[sigconf,nonacm]{acmart}

\usepackage{graphicx} %
  \graphicspath{{./figs/}}
  \DeclareGraphicsExtensions{.pdf,.tex,.png}
\usepackage{siunitx}

\usepackage{xcolor}
\usepackage{listings}

\usepackage{tikz}
\usetikzlibrary{shapes,arrows}
\usetikzlibrary{calc}
\usetikzlibrary{fit,positioning}
\usetikzlibrary{patterns}
\usetikzlibrary{backgrounds}
\usetikzlibrary{decorations.pathreplacing}

\lstdefinestyle{myShell}{
backgroundcolor=\color{black!10!white},
basicstyle=\ttfamily\scriptsize\linespread{-0.5},
breakatwhitespace=false,         
breaklines=true,                 
captionpos=b,                    
keepspaces=true,                 
numbersep=5pt,                  
showspaces=false,                
showstringspaces=false,
showtabs=false,                  
tabsize=2
}

\begin{document}

\title{The Kieker Observability Framework Version~~2}
\author{Shinhyung Yang}
\email{shinhyung.yang@email.uni-kiel.de}
\orcid{0000-0002-8997-9942}
\affiliation{%
  \institution{Kiel University}
  \city{Kiel}
  \state{Schleswig-Holstein}
  \country{Germany}
}
\author{David Georg Reichelt}
\email{d.g.reichelt@lancaster.ac.uk}
\orcid{0000-0002-1772-1416}
\affiliation{%
  \institution{Lancaster University Leipzig \&%
  ~Universit{\"a}t Leipzig}
  \city{Leipzig}
  \state{Saxony}
  \country{Germany}
}
\author{Reiner Jung}
\email{reiner.jung@email.uni-kiel.de}
\orcid{0000-0002-5464-8561}
\affiliation{%
  \institution{Kiel University}
  \city{Kiel}
  \state{Schleswig-Holstein}
  \country{Germany}
}
\author{Marcel Hansson}
\email{marcel.hansson@uni-hamburg.de}
\orcid{0009-0000-6524-037X}
\affiliation{%
  \institution{University of Hamburg}
  \city{Hamburg}
  \country{Germany}
}
\author{Wilhelm Hasselbring}
\email{hasselbring@email.uni-kiel.de}
\orcid{0000-0001-6625-4335}
\affiliation{%
  \institution{Kiel University}
  \city{Kiel}
  \state{Schleswig-Holstein}
  \country{Germany}
}

\thanks{%
This is the author’s version of the work for an arXiv submission. The
definitive Version of Record can be found at
\url{https://doi.org/10.1145/3680256.3721972}, published in Companion of the
16th ACM/SPEC International Conference on Performance Engineering (ICPE
Companion ’25), May 5--9, 2025, Toronto, ON, Canada.%
}

\begin{abstract}
Observability of a software system aims at allowing its engineers and
operators to keep the system robust and highly available. With this paper, we
present the Kieker Observability Framework Version~2, the successor of the
Kieker Monitoring Framework. 
  
In this tool artifact paper, we do not just present the Kieker framework, but
also a demonstration of its application to the Tea\-Store benchmark,
integrated with the visual analytics tool ExplorViz. This demo is provided
both as an online service and as an artifact to deploy it yourself.
\end{abstract}



\maketitle

\section{Introduction}

Observability is a fundamental property of a software system that should be
considered during system design~\cite{Observability1960}. Using telemetry data
allows various stakeholders to understand the system and answer key questions.
Monitoring and testing are integral parts of observability, and both benefit
from and contribute to it. Observability of a system has become significant as
the architecture of software systems is increasingly containerized with distributed
microservices. These requirements have conceived observability tools that
provide insights into the internal behavior of software systems using traces,
metrics, and logs~\cite{Gatev2021}.

In this tool artifact paper, we introduce the Kieker Observability Framework
Version 2, the successor of the Kieker Monitoring Framework. We demonstrate
Kieker's observability, analysis, and visualization with our
TeaStore-Kieker-ExplorViz demo~\cite{Reichelt2024KiekerOTelExplorviz}, which
includes our initial effort for the interoperability between Kieker and
OpenTelemetry~\cite{GomezBlanco2023}.

The new contribution of this tool artifact paper is not just the Kieker
framework, but also a demonstration of its application to the TeaStore
benchmark~\cite{Teastore2018}, integrated with the visual analytics tool
ExplorViz~\cite{ExplorViz2020}. This demo is provided both as an online service
and as an artifact to deploy it yourself.

Section~\ref{s-Impact} summarizes Kieker's impact, so far. Observability
Engineering is introduced in Section~\ref{ssec:observability}, before
instrumentation and data collection for observability with Kieker is presented
in Section~\ref{s-Instrumentation}. The tool artifact is described in
Section~\ref{sec:Artifact}. Section~\ref{sec:Related} discusses some related
work, before the paper is summarized in Section~\ref{sec:Summary} with an
outlook to future work.

\section{Impact of Kieker}\label{s-Impact}

The Kieker Observability Framework started as a monitoring framework in
2006~\cite{Kieker2020}, supporting application performance monitoring and
analysis. Kieker enables observing software systems with Java agents, which
produce metrics, logs, and trace data. Using the trace data, Kieker allows for
reverse engineering and visualization of the software architecture. The Kieker
developers have been continuously monitoring its performance overhead with the
MooBench monitoring overhead microbenchmark~\cite{MooBench2015, Reichelt2021}.

Kieker allows for the analysis of observed data. The new Kieker Version 2
incorporates the TeeTime pipe-and-filter framework~\cite{TeeTime2017}, which has
restructured the analysis pipeline. It provides an intuitive way for Kieker
users to build new analysis applications.

Research areas in which Kieker was successfully employed are performance
analysis~\cite{SIPEW2008Rohr,SSP2015Zirkelbach,Cito2018performancehat,Reichelt2016,PeASS2019},
benchmarking streaming engines~\cite{Yang2018}, anomaly detection~\cite{CSMR2009,ICAC2011}, online
capacity management~\cite{WUP2009,ECSA2011Massow}, analysis of software
structure and
metrics~\cite{Jin2021,Qu2015exploring,schnoor_comparing_2020,Liu2022,Cao2022,Andrade2023},
software architecture
reconstruction~\cite{Dabrowski2012OnArchitectureWarehousesAndSoftwareIntelligence},
test case prioritization \cite{chi2020}, and extraction of load
profiles~\cite{ValueTools2014,WESSBAS2016}. \citet{MartinezSaucedo2025} recently
identified Kieker as the most cited tool to assist with the migration of
monolithic systems to microservices.

Several Kieker-related research projects have been and are being conducted over
the years. Examples are
diagnoseIT (Expert-Guided Automatic Diagnosis of Performance Problems)~\cite{HegerHOSW16}, 
DynaMod  (Dynamic Analysis for Model-Driven Software
Modernization)~\cite{MDSM2011Dynamod}, and OceanDSL (Architecture analysis,
interactive visualization) \cite{OceanDSL2021}. As reported by
\citet{KiekerTransfer2015}, Kieker was also employed in several industrial
collaborations and technology transfer projects. More information on Kieker's
impact may be found in \citet{Kieker2020}.

Kieker is included in the SPEC Research Group's repository of peer-reviewed
tools for quantitative system evaluation and
analysis.\footnote{\url{https://research.spec.org/tools/overview/}} The review
process and the final acceptance of Kieker for this tool repository triggered
manifold activities in the Kieker project and helped to further improve Kieker's
product quality (both code and documentation); Kieker's visibility increased
considerably, with \textbf{480} citations to the ICPE~2012 Kieker tool demo
paper \cite{Kieker2012}, so
far.\footnote{\url{https://scholar.google.com/scholar?cluster=11724904481060384258}
(Jan.\ 15, 2025)}

\section{Observability Engineering\label{ssec:observability}}

The term observability has been introduced in the context of control
theory~\cite{Observability1960}. In this context, observability is a measure of
how well the internal states of a system can be inferred from knowledge of its
external outputs. For software systems, this means how well the internal state
of a software system can be inferred from its monitoring data. Observability
Engineering~\cite{majors_observability_2022} focuses on designing, building, and
maintaining systems that enable us to understand the internal state of software
systems based on monitoring data. It is a crucial part of software engineering,
especially in the context of complex distributed systems such as microservices
and cloud-native architectures. Figure~\ref{fig:ThreePillarsOfObservability}
(introduced by Peter
Bourgon\footnote{\url{https://peter.bourgon.org/blog/2017/02/21/metrics-tracing-and-logging.html}})
presents the three pillars of observability: metrics, logging, and
tracing~\cite{Gatev2021}:

\begin{figure}[bht]
  \includegraphics[width=\linewidth]{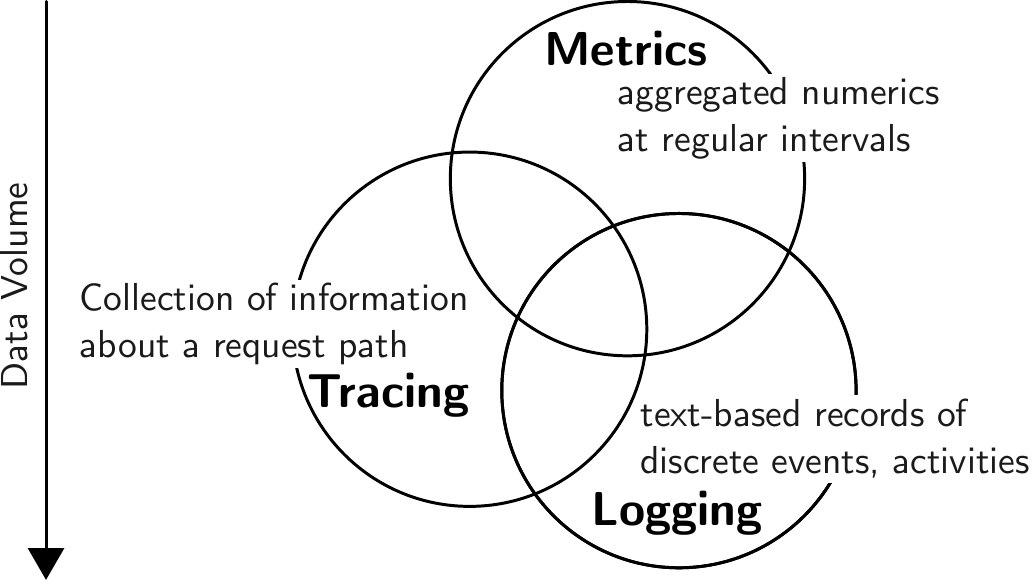}
  \caption[ThreePillarsOfObservability]%
  {Visual illustration of metrics, logging, and tracing. They are ordered
  vertically with a higher data volume at the bottom.}%
  \Description[ThreePillarsOfObservability]{The Three Pillars of Observability}
  \label{fig:ThreePillarsOfObservability}
\end{figure}

\begin{itemize}
\item Metrics are aggregated numerical data representing the performance and
behavior of systems (e.g., CPU usage, response time, error rates), which are
collected at regular intervals.
\item Logging is the collection of text-based records of discrete events or
activities within a system (e.g., status and debugging messages).
\item Tracing is the collection of information about the path a request takes as
it moves through a distributed system. Traces are aggregations of highly
structured log events, which are called \textit{spans}~\cite{GomezBlanco2023}.
\end{itemize}
Observability is not a substitute for monitoring, nor does it obviate the need
for monitoring; they are complementary. Observability is a fundamental property
of a software system that must be intentionally considered during system
design~\cite{majors_observability_2022}. It enables different stakeholders to
understand the system and answer key questions by utilizing telemetry data.
Monitoring and testing are integral parts of observability, and both benefit
from and contribute to it.

\section{Instrumentation and Data Collection with Kieker}\label{s-Instrumentation}

The three pillars of observability~\cite{Gatev2021} are addressed with Kieker as
follows:
\begin{itemize}
\item Metrics are sampled using Kieker's periodic sampler. It allows for the
incorporation of different samplers to collect metrics. Kieker currently
supports sampling the JVM garbage collector and sampling CPU and memory
information.
\item Logs are supported by Kieker as a generic way to handle event-centric
messages. 
\item Traces are recorded by Kieker's tracing agents. Kieker features
distributed tracing that records the internal behaviors of the target system in
the scopes of components in the deployed microservices. Kieker traces consist of
records that are spans.
\end{itemize}
Kieker enables observability of a software system by collecting data with Kieker
agents. Furthermore, Kieker is designed to analyze and visualize the observed
software system. Below, we take
a look at instrumenting software applications with Kieker
(Section~\ref{ss-Instrumentation}), and on benchmarking the incurred overhead
(Section~\ref{ss-Overhead}).

\subsection{Instrumentation}\label{ss-Instrumentation}

Observability data can be obtained either using instrumentation of code or
sampling of existing data collection interfaces, e.g., JMX beans or tracepoints
of the Linux kernel. Kieker focuses on instrumentation of application code.
This can be done either automatically or via manual source code instrumentation.
With Java, the automated instrumentation is performed by the
``\texttt{-javaagent}'' interface of the JVM, which allows developers to change
the bytecode during application loading. Kieker supports instrumentation via
AspectJ, ByteBuddy, DiSL, and Javassist \cite{Reichelt2024}. Kieker also
supports automated instrumentation of existing source code
\cite{reichelt2023towards}, if the Java source code is available. Additionally,
Kieker supports automated injection into servlets and Spring applications, and
sampling CPU metrics using various OSHI~(Operating System and Hardware
Information library) samplers. Besides its Java instrumentation, Kieker allows
to instrument C, Fortran and Python code
\cite{jung2021instrumenting,simonov2023instrumenting}.

\subsection{Benchmarking Observability Overhead}\label{ss-Overhead}

The instrumentation for observability imposes overhead, which should be
minimized in production environments. Kieker's overhead has been continuously
measured since
2015\footnote{\url{https://kieker-monitoring.net/performance-benchmarks/}} using
the Moo\-Bench microbenchmark \cite{MooBench2015}. It measures both the overall
overhead of Kieker tracing and the overhead that is incurred by different
factors, which are the instrumentation itself, the measurement, and the final
serialization of data. Since the measurement of overhead in the JVM requires
coping with its non-determinism, the measurement requires the repetition of the
measured workload. Therefore, MooBench repeats a method call for a specified
iteration count. This method recursively calls itself for a given recursion
depth $d$, and waits for a specified duration $t$ in the leaf node. 

MooBench also measures the overhead of the tracing tools OpenTelemetry and
inspectIT. Studies show that the overhead incurred by these tools is
significantly higher than the overhead of Kieker's
instrumentation~\cite{Reichelt2021}. This is partially achieved by the complete
decoupling of the application thread and the log writer thread. The application
thread executes asynchronously immediately after a Kieker record has been
written into the queue \cite{MSEPT2012,strubel2016refactoring}.

\section{The Tool Artifact\label{sec:Artifact}}

\begin{figure*}[ht]
  \includegraphics[width=\linewidth]{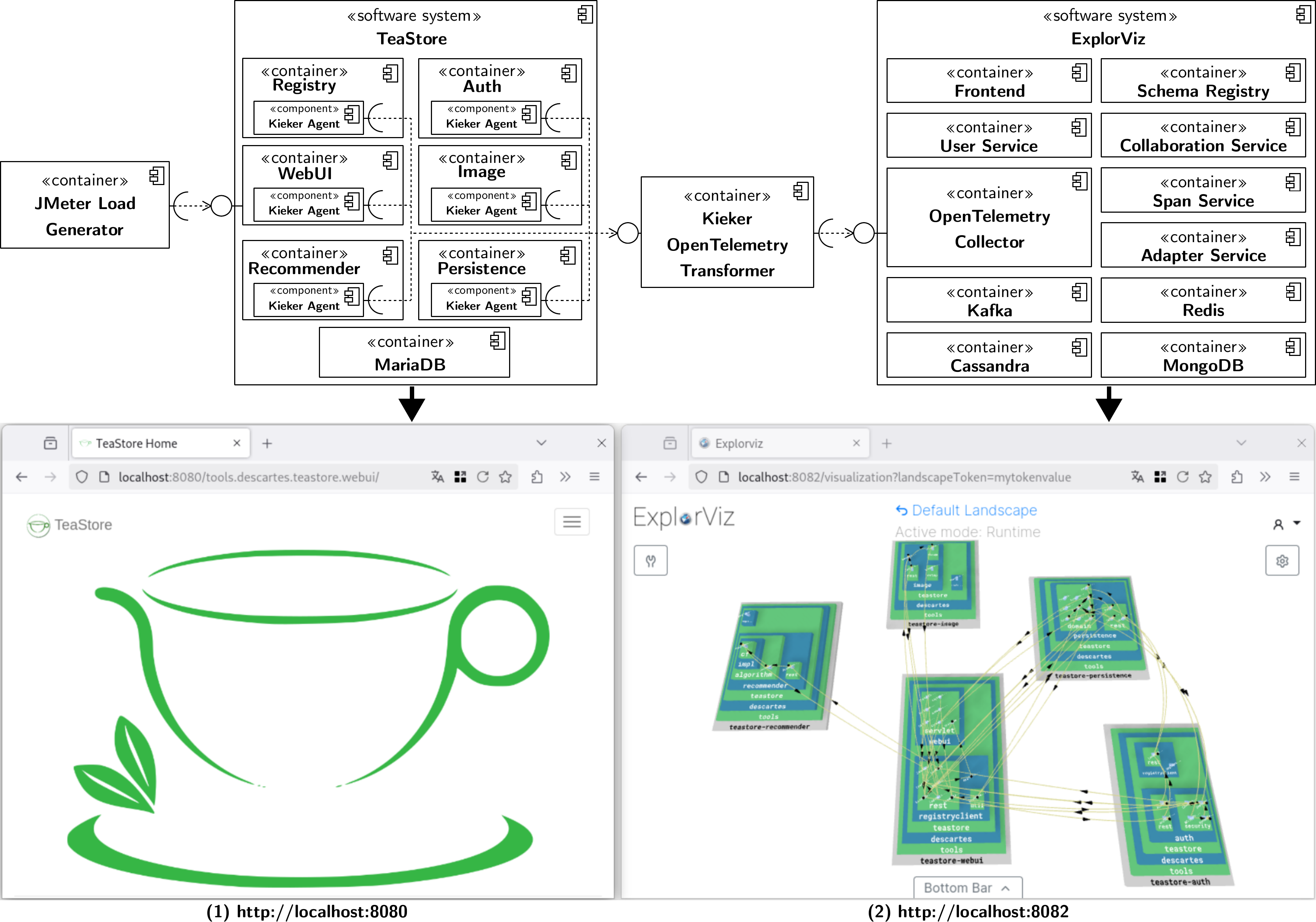}
  \caption{Architecture of the tool artifact. The provided Docker compose file
launches all the containers in the component diagram, and it is accessible with
two web servers on the Ports 8080~(1) and 8082~(2).}
  \Description[The Architecture of the Demo]{The Architecture of the Demo}
\label{fig:DemoArchitecture}
\end{figure*}

The architecture of our tool artifact is illustrated in
Figure~\ref{fig:DemoArchitecture}. Our tool artifact consists of four software
systems (top-level components), which are available as an online services
(non-installation) and an offline docker compose file. 

The TeaStore software systems consists of seven microservices, each deployed in
its own Docker container. The TeaStore microservices, except the database
(MariaDB), are instrumented via Kieker agents. The ExplorViz software system
consists of eleven microservices, again each deployed in its own Docker
container. The connection between the TeaStore system and the ExplorViz system
is accomplished via the embedded Kieker agents that send the monitoring data to
the Kieker OpenTelemetry Transformer, which forwards it after transformation to
the OpenTelemetry
Collector.\footnote{\url{https://opentelemetry.io/docs/collector/}} The
OpenTelemetry Collector is part of the ExplorViz system, as one of its
microservices.

First, JMeter load tests the TeaStore microservices via a configurable number of
requests. The Kieker agents in the TeaStore microservices then send the traces
to the Kieker OpenTelemetry Transformer. The Kieker OpenTelemetry Transformer
consists of several TeeTime stages~\cite{TeeTime2017}, which translate the
received Kieker traces into OpenTelemetry spans. The ExplorViz tool receives the
translated spans and renders the dynamic visualization of the TeaStore
architecture.

The online service has two URLs, one for the TeaStore Web
UI,\footnote{\url{https://teastore.sustainkieker.kieker-monitoring.net/}}
and another URL for the ExplorViz Web
UI.\footnote{\url{https://explorviz.sustainkieker.kieker-monitoring.net/}}
Kiel University hosts the online service, which is available on the
SustainKieker homepage.

\paragraph{Installation}

Our tool artifact runs with a Docker compose script file. We use it to
(1)~retrieve Docker images from DockerHub, or (2)~build the Docker images
locally for archival purposes. The script simplifies the deployment of the
TeaStore and ExplorViz software systems, JMeter, and the Kieker OpenTelemetry
Transformer. We tested the out-of-the-box experience on all major platforms,
Linux, macOS, and Windows. For the open-source access, we use a GitHub
repository\footnote{\url{https://github.com/kieker-monitoring/tool-artifact}} to
provide our tool artifact. A Zenodo upload is available to download and run the current
snapshot of the tool artifact.\footnote{\url{https://doi.org/10.5281/zenodo.14989908}}

The tool artifact requires installation of Docker and Git (optional) on
all platforms. The installation and launching of our tool artifact follows
Listing~\ref{lst:installation}.
\begin{lstlisting}[label=lst:installation,style=myShell,caption={Installing and launching the tool artifact}]
  git clone https://github.com/kieker-monitoring/tool-artifact
  cd tool-artifact
  docker compose up -d
\end{lstlisting}

After the tool artifact launches, two web servers become available: the TeaStore
WebUI on the TCP port~8080 (Figure~\ref{fig:DemoArchitecture}.(1))  and the
ExplorViz frontend on the TCP port~8082 (Figure~\ref{fig:DemoArchitecture}.(2)).

ExplorViz \cite{ExplorViz2020} is an open-source research visualization tools,
which uses dynamic analysis techniques to provide a live trace visualization of
software landscapes, in our case the microservices of the TeaStore. ExplorViz
targets system and program comprehension for software landscapes or single
applications while still providing details on the communication within an
application. It utilizes the 3D city metaphor combined with an interactive
concept of showing only details that are in focus of the analysis.

The JMeter load driver initially launches a TestPlan that generates \num{6400}
requests on the TeaStore. This way, ExplorViz can visualize those parts of the
TeaStore that were called for this load. Afterwards, only the interaction of a
user with the TeaStore Web interface is observed and visualized with ExplorViz.

\paragraph{Documentation}
Comprehensive documentation, including an introduction and a quick start guide
for Kieker Version 2 is available
online.\footnote{\url{https://kieker-monitoring.readthedocs.io/en/latest/index.html}}
In addition, a (JavaDoc) API
documentation\footnote{\url{https://api.kieker-monitoring.org/2.0.2/}} provides
insights for developers to further expand or work with the Kieker source files.

\section{Related Work}\label{sec:Related}

Related work exists in the field of \textit{observability tools} and \textit{software visualization}.

Concerning \textit{observability tools}, a variety of alternatives to Kieker
exists~\cite{Janes2023tracers}. Example proprietary tools are DataDog\footnote{\url{https://docs.datadoghq.com/agent}} and the DynaTrace One
Agent.\footnote{\url{https://www.dynatrace.com/platform/oneagent/}}
Example open source tools, are Jaeger\footnote{\url{https://www.jaegertracing.io/}} and
Elastic APM.\footnote{\url{https://www.elastic.co/observability/}}
For these observability tools integrations
with OpenTelemetry exist, as presented in our work. Therefore, they could
also be used for TeaStore instrumentation and the date export to ExplorViz; however, there is currently no available implementation.

\citet{yang2022cloudprofiler} present Cloudprofiler, a tool for profiling of
systems that are processing streaming workloads. They focus on the
synchronization of multiple virtual machines and compress the logs to reduce
I/O usage. They benchmark Cloudprofiler using the Yahoo streaming benchmark and
measure an overhead of~2.2\,\%. The MooBench benchmark with Cloudprofiler finds
that the overhead of Cloudprofiler is close to the overhead of
Kieker~\cite{yang2024evaluating}. An integration of Cloudprofiler and the
OpenTelemetry standard does not exist currently, therefore, it could not be used
to instrument the TeaStore and export observability data to ExplorViz.

Concerning \textit{visualization}, different visualizations of Kieker traces exist. Müller and Fischer~\cite{muller2019graph} provide an
alternative visualization of Kieker traces based on jQAssist. Although
they also visualize call trees, their visualization is less flexible than the
ExplorViz UI. Alternative approaches use static software analysis
data~\cite{hoff2022softwarearchitecture}. While they can also
visualize the component structure of an application, they cannot
present runtime information such as the number of calls between the components, since
these data need to be obtained by instrumentation.

\section{Summary and Outlook}\label{sec:Summary}

The application of the Kieker observability framework to the TeaStore benchmark,
integrated with the visual analytics tool ExplorViz is presented in this paper.
This tool artifact is provided both as an online service and as an artifact to
deploy it yourself.

With our current SustainKieker
project,\footnote{\url{https://sustainkieker.kieker-monitoring.net/}} we intend
to further sustain Kieker as a reusable, high-quality observability framework
that will be employed in a wider variety of research areas and used by a larger
community. The online version of the tool artifact, as presented in this paper,
also serves as a demo application to use Kieker for observability.

Several demo applications to be developed in SustainKieker will serve as basis
for new interactive tutorial examples for Kieker. Besides the online manual and
step-by-step instructions in tutorials, we aim to produce screencasts based on
the demos illustrating core features of Kieker. We will also empirically
evaluate the effect of our new tutorials with the embedded online demos.

\section*{Acknowledgment}
This research is funded by the Deutsche Forschungsgemeinschaft (DFG -- German
Research Foundation), grant no.~528713834.

\bibliographystyle{ACM-Reference-Format}
\balance
\bibliography{references.bib}

\end{document}